\documentclass[12pt]{article}
\usepackage{graphicx,epsfig,cite,amsmath}

\newcommand{\be}{\begin{equation}}
\newcommand{\ee}{\end{equation}}
\newcommand{\ba}{\begin{eqnarray}}
\newcommand{\ea}{\end{eqnarray}}

\newcommand{\chpt}{$\chi$PT}
\newcommand{\cO}{{\cal O}}
\newcommand{\cJ}{{\cal J}}

\begin{document}
\begin{titlepage}
\begin{flushright}
{CAFPE-98/08}\\  {FTUV/08-1004} \\ IFIC/08-44 \\ {UG-FT-228/08}
\end{flushright}
\vspace{2cm}
\begin{center}
{\large\bf Determination of the Chiral Couplings $\mathbf{L_{10}}$
and $\mathbf{C_{87}}$\\[10pt]  from Semileptonic\ {\Large $\mathbf{\tau}$}\ Decays}
\vfill
{\bf Mart\'{\i}n Gonz\'alez-Alonso$^a$, Antonio Pich$^a$ and
Joaquim Prades$^b$}\\[0.5cm]
${}^a$
Departament de F\'{\i}sica Te\`orica and IFIC, Universitat
de Val\`encia-CSIC,\\
Apt. Correus 22085, E-46071 Val\`encia, Spain\\
${}^b$ CAFPE and Departamento de F\'{\i}sica Te\'orica y del Cosmos,\\
Universidad de Granada, Campus de Fuente Nueva, E-18002 Granada,
Spain\\[0.5cm]
\end{center}
\vfill
\begin{abstract}
Using recent precise hadronic $\tau$-decay data on the $V-A$ spectral function,
and general properties of QCD such as analyticity, the operator product expansion
and chiral perturbation theory, we  get accurate
values for the QCD chiral order parameters $L_{10}^r(M_\rho)$
and $C_{87}^r(M_\rho)$. These two low-energy constants
appear at order $p^4$ and $p^6$, respectively,
in the chiral perturbation theory expansion of the $V-A$ correlator.
At order $p^4$ we obtain
$L_{10}^r(M_\rho) = -(5.22\pm 0.06)\cdot 10^{-3}$.
Including in the analysis the two-loop (order $p^6$) contributions, we get
$L_{10}^r(M_\rho) = -(4.06\pm 0.39)\cdot 10^{-3}$ and
$C_{87}^r(M_\rho) = (4.89\pm 0.19)\cdot 10^{-3}\;\mathrm{GeV}^{-2}$.
In the SU(2) chiral effective theory, the corresponding low-energy coupling
takes the value $\overline l_5 = 13.30 \pm 0.11
$ at order $p^4$, and $\overline l_5 = 12.24 \pm 0.21$ at order $p^6$.
\end{abstract}
\vfill

\end{titlepage}

\section{Introduction}
The precise hadronic $\tau$-decay data provided in
refs.~\cite{ALEPH05,ALEPH98,ALEPH97,OPAL99,CLEO95,Strange_data}
are a very important source of information, both on perturbative and
non-perturbative QCD parameters.
The theoretical analysis of the inclusive $\tau$ decay width into hadrons
allows to perform an accurate determination of the QCD coupling
$\alpha_s(M_\tau)$ \cite{alphas,LDP:92,DHZ06,new08,review},
which becomes the most precise determination of $\alpha_s(M_Z)$
after QCD running.
In this case, non-perturbative QCD effects
parametrised by  power corrections are strongly suppressed.
Another example of the use of hadronic $\tau$-decay data
is the study of SU(3)--breaking corrections to the
strangeness-changing two-point functions \cite{su3,CKP98,KKP01,MALT,Vus}.
The separate measurement of the $|\Delta S|=0$ and $|\Delta S|=1$ tau decay widths
provides accurate determinations of fundamental parameters
of the Standard Model, such as the strange quark mass and the
Cabibbo-Kobayashi-Maskawa quark-mixing $|V_{us}|$ \cite{Vus}.

Very important phenomenological hadronic matrix elements and
non-perturbative QCD quantities can also be obtained from $\tau$-decay data.
Of special interest is the difference of the vector and axial-vector spectral functions,
because in the chiral limit the corresponding $V-A$ correlator is exactly zero in perturbation theory.
The $\tau$-decay measurement of the $V-A$  spectral function has been used to perform \cite{DG:94,DHG98,NAR01} phenomenological tests
of the so-called Weinberg sum rules (WSRs) \cite{WSR}, to compute the electromagnetic mass difference between the charged
and neutral pions \cite{DHG98}, and to determine several QCD vacuum condensates \cite{DS07,CGM03}. From the same spectral function one
can also determine the $\Delta I=3/2$ contribution of
the $\Delta S=1$ four-quark operators $Q_7$ and $Q_8$ to $\varepsilon_K'/\varepsilon_K$, in the chiral limit \cite{Q7Q8}.

Using chiral perturbation theory (\chpt) \cite{WEI79,GL84,GL85},
the hadronic $\tau$-decay data can also
be related to order parameters of the spontaneous chiral
symmetry breaking (S$\chi$SB) of QCD \cite{KdR94}.
\chpt\ is the effective field theory of QCD at very low energies;
it describes the S$\chi$B Nambu-Goldstone boson physics
through  an expansion in external momenta and quark masses. The
coefficients of that expansion 
are related to order parameters of S$\chi$SB.
At lowest order (LO), i.e. ${\cal O}(p^2)$, all low-energy observables
are described in terms of the pion decay constant $f_\pi \simeq 92.4$ MeV and
the light quark condensate.
At next-to-leading order (NLO), ${\cal O}(p^4)$, the SU(3) \chpt\ Lagrangian contains 12
low-energy constants (LECs), $L_{i=1,\cdots,10}$ and $H_{1,2}$
 \cite{GL85}.
At ${\cal O}(p^6)$, 90 (23) additional parameters $C_{i=1,\cdots,90}$
appear in the
even (odd) intrinsic parity sector \cite{p6}.
These LECs are not fixed by symmetry requirements alone and have to be determined
phenomenologically or using non-perturbative techniques.
The ${\cal O}(p^4)$ $L_i$ couplings have been determined in the past to an acceptable accuracy;
a recent compilation can be found in ref.~\cite{ECK07}.
Much less well determined are the ${\cal O}(p^6)$ couplings $C_i$.

There has been a lot of recent activity to determine the chiral LECs
from theory, using as much as possible QCD information
\cite{MOU97,KN01,RPP03,BGL03,CEE04,CEE05,KM06,RSP07,MP08,PRS08}.
This strong effort is motivated by the precision required in present phenomenological applications,
which makes necessary to include corrections of ${\cal O}(p^6)$. The huge number of unknown couplings is the major source of theoretical uncertainty.

In this paper we present an accurate
determination of the \chpt\ couplings $L_{10}$ and $C_{87}$,
using the most recent experimental data on hadronic $\tau$ decays \cite{ALEPH05}.
Previous work on $L_{10}$ using $\tau$-decay data can be found in
refs.~\cite{DHG98,NAR01,DS07,DS04}. Our analysis is the first one which includes
the known two-loop \chpt\ contributions and, therefore, provides also
the $\cO(p^6)$ coupling $C_{87}$.

%First, we discus the theoretical input needed and  we analyse the results
%of using the recent hadronic $V-A$ tau data.
% We also compare  these results with other
%recent analytic results and hadronic tau data determinations.
%Finally, we give our results and conclusions.

\section{Theoretical Framework}

The basic objects of the theoretical analysis are the two-point
correlation functions of the vector and axial-vector quark currents, defined as follows:
\ba
\label{eq:two}
\Pi^{\mu\nu}_{ij,\cJ}(q)
&\equiv & i \int \mathrm{d}^4 x \;  \mathrm{e}^{i q x} \,
\langle 0 | T \left( \cJ_{ij}^\mu(x) \cJ_{ij}^\nu(0)^\dagger \right) | 0 \rangle
\nonumber \\ &=& (-g^{\mu\nu} q^2 + q^\mu q^\nu ) \, \Pi^{(1)}_{ij,\cJ}(q^2)
+ q^\mu q^\nu\, \Pi^{(0)}_{ij,\cJ}(q^2) \, .
\nonumber \\
% \nonumber \\
%\Pi^{\mu\nu}_{ij,A}(q)
%&\equiv& i \int {\rm d}^4 x \, e^{i q x} \,
%\langle 0 | T \left( A_{ij}^\mu(x) A_{ij}^\nu(0)^\dagger \right) | 0 \rangle
%\nonumber \\ &=& (-g^{\mu\nu} q^2 + q^\mu q^\nu ) \, \Pi^{(1)}_{ij,A}(q^2)
%+ q^\mu q^\nu \Pi^{(0)}_{ij,A}(q^2) \, .
\ea
Here, we just need  the non-strange correlators, i.e. $\cJ_{ij}^\mu(x)$ denotes the
Cabibbo-allowed vector or axial-vector currents,
 $V_{ud}^\mu(x)=\overline{u} \gamma^\mu d$ and
$A_{ud}^\mu=\overline{u} \gamma^\mu \gamma_5 d$.
Moreover, our analysis will concentrate in the difference
\ba
\Pi(s)\, &\equiv&\, \Pi_{ud,V-A}^{(0+1)}(s)\, =\,\Pi_{ud,V}^{(0+1)}(s)-\Pi_{ud,A}^{(0+1)}(s) \nonumber \\
\,&\equiv&\, 
\frac{2 f_\pi^2}{s-m_\pi^2} + \overline{\Pi}(s)\, ,
\ea
where we have made explicit the contribution of the pion pole to the longitudinal axial-vector
two-point function. We will work in the isospin limit $m_u=m_d$ where  $\Pi^{(0)}_{ud,V}(q^2)=0$.

%%%%%%%%%%%%%%%%%%% FIGURE %%%%%%%%%%%%%%%%%%%%%
\begin{figure}[htb]
\centering
\includegraphics[width=0.4\textwidth]{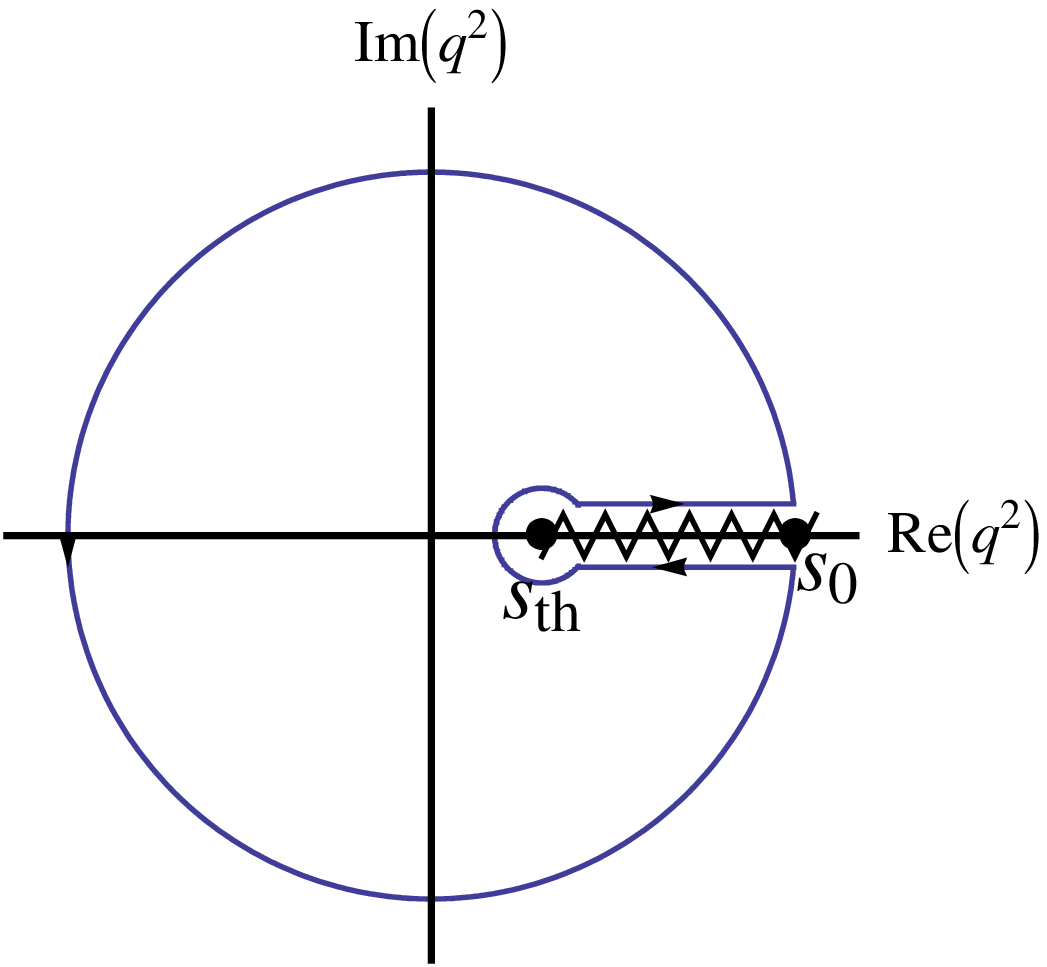}
\caption[]{Analytic structure of $\overline{\Pi}(s)$.}
% The cut is represented by a zig-zag line.}
% Notice that the weight functions in (\ref{eq:two}) have a pole at the origin when $n$ is negative.
\label{fig:circuit}
\end{figure}
%%%%%%%%%%%%%% END FIGURE %%%%%%%%%%%%%%%%%%%%%%

The correlator $\overline{\Pi}(s)$ is analytic in the entire complex $s$-plane, except for a cut on the
positive real axis which starts at the threshold $s_{\mathrm{th}}=4 m_\pi^2$.
Applying Cauchy's theorem to the circuit in Fig.~\ref{fig:circuit}, one gets the exact relation:
\ba
\label{eq:sumrules}
&& \int^{s_0}_{s_{\rm th}} \mathrm{d}s\; s^n \, \frac{1}{\pi} \, {\rm Im} \, \Pi(s)
\, +\,  \frac{1}{2 \pi i} \, \oint_{|s|=s_0} \mathrm{d}s\; s^n \,\Pi(s) \nonumber \\
& =& 2 f_\pi^2\, m_\pi^{2n}
+ \mathrm{Res} \left[ s^n \, \Pi(s), s=0\right] .\quad
\ea
For non-negative values of the integer power $n$, the pion pole is the only singularity within the contour
and one gets the so-called finite energy sum rules (FESR), widely used in the literature.
When $n$ takes negative values, the weight factor $s^n$ introduces a pole at the origin which
gives rise to the additional contribution in the r.h.s. of the equation, given by the residue of
$s^n \Pi(s)$ at $s=0$.

In the chiral limit ($m_u=m_d=0$) the correlator $\Pi(s)$ vanishes identically to all orders
in perturbation theory. For large enough Euclidean values of $s=-Q^2$ its operator product expansion (OPE),
$\Pi(Q^2) = \sum_k C_{2k}^{V-A}/Q^{2k}$,
contains only power-suppressed contributions from dimension $d=2k$ operators,
starting at $d=6$. The nonzero up and down quark masses induce tiny corrections with dimensions
two and four, which are negligible at high values of $Q^2$.
Therefore, with $n\ge 0$ and $s_0$ large enough so that the OPE can be applied in the entire circle
$s=s_0$, the integral over the spectral function from $s_{\mathrm{th}}$ to $s_0$ is equal to
the pion pole term $2 f_\pi^2\, m_\pi^{2n}$ plus the OPE contribution
$(-1)^n C_{2(n+1)}^{V-A}$
generated by the integration along the circle.
For $n=0$ and $n=1$, $C_{2(n+1)}^{V-A}$ is zero in the chiral limit and
one gets the celebrated first and second WSRs \cite{WSR}, respectively.

For negative values of $n\equiv -m<0$, the OPE does not give any contribution to the integration along
the circle $s=s_0$. One gets then:
\ba
\label{eq:nSR}
&& \int^{s_0}_{s_{\rm th}} \frac{\mathrm{d}s}{s^m} \; \frac{1}{\pi} \, {\rm Im} \, \Pi(s)
\; =\; \frac{2 f_\pi^2}{m_\pi^{2m}} \, 
+\,\frac{1}{(m-1)!}\,\Pi^{(m-1)}(0) \nonumber \\
\, &=&\, \frac{1}{(m-1)!}\,\overline{\Pi}^{(m-1)}(0)\, ,
\ea
where $\overline{\Pi}^{(m-1)}(0)$ denotes the $(m-1)$th derivative of $\overline{\Pi}(s)$ at
$s=0$.
The interest of this relation stems from the fact that at low values of $s$ the correlator can be rigourously
calculated within \chpt. At present $\Pi(s)$ is known
to $\cO(p^6)$ \cite{ABT00},
in terms of the LECs that we want to determine. The choices $m=1$ and $m=2$
allow then us to relate the spectral function measured
in $\tau$ decays with the theoretical expressions of $\overline{\Pi}(0)$ and $\overline{\Pi}\,{}'(0)$,
which can be derived from the results obtained in ref.~\cite{ABT00}:
\ba
\label{L10-p6}
L_{10}^{\mathrm{eff}} &\!\equiv &\! -\frac{1}{8}\, \overline{\Pi}(0)
\nonumber\\ &\! = &\!
L_{10}^r(\mu) \, +\,
\frac{1}{128 \,\pi^2} \left[1- \log{\left(\frac{\mu^2}{m_\pi^2}\right)}\,
+ \,\frac{1}{3}\,\log{\left(\frac{m_K^2}{m_\pi^2}\right)} \right]
\nonumber\\ &\! +&\!
4 m_\pi^2 \left( C_{61}^r - C_{12}^r - C_{80}^r\right)\! (\mu)
 \, \nonumber \\ &+& \, 4 \left( 2 m_K^2 + m_\pi^2 \right)\,
\left( C_{62}^r - C_{13}^r - C_{81}^r \right)\! (\mu)
\nonumber\\[7pt] &\! - &\!
2\, \left( 2 \mu_\pi + \mu_K \right) \,\left( L_9^r + 2 L_{10}^r\right)\! (\mu)\, \nonumber \\
&+&\,  G_{2L}(\mu,s\!=\!0) \, +\, {\cal O} (p^8) \, ,
%%%-\,\frac{1}{8\, f_\pi^2} \, G_{2L}(\mu) \, ,
\\[10pt]\label{C87-p6}
C_{87}^{\mathrm{eff}} &\!\equiv &\! \frac{1}{16}\, \overline{\Pi}\,{}'(0)
\nonumber\\ &\! = &\!
C_{87}^r(\mu) \! + \!
\frac{1}{7680\, \pi^2} \left( \frac{1}{m_K^2} + \frac{2}{m_\pi^2} \right)
\nonumber \\ &-& \,
\frac{1}{64 \,\pi^2 f_\pi^2} \left[1- \log{\left(\frac{\mu^2}{m_\pi^2}\right)}\,
+ \,\frac{1}{3}\,\log{\left(\frac{m_K^2}{m_\pi^2}\right)} \right] \, L_9^r(\mu)
\nonumber \\ \, &-&
\frac{1}{2}\, G'_{2L}(\mu,s\!=\!0)  \, +\, {\cal O} (p^8)\, ,
%%% \frac{1}{16\, f_\pi^2}\, G'_{2L}(\mu)  \, + {\cal O} (p^8) \, ,
\ea
where $\mu_i= m_i^2 \log(m_i/\mu)/(16 \pi^2 f_\pi^2)$.

To a first approximation the effective parameters $L_{10}^{\mathrm{eff}}$
and $C_{87}^{\mathrm{eff}}$ correspond to the LECs $L_{10}^r(\mu)$ and $C_{87}^r(\mu)$, respectively.
At $\cO(p^4)$, the only relevant correction is given by
the logarithmic terms in the second line
of (\ref{L10-p6}), which cancel the \chpt\ renormalization
scale dependence of $L_{10}^r(\mu)$;
these contributions are suppressed by one power of $1/N_C$
with respect to $L_{10}^r(\mu)$, where
$N_C$ is the number of quark colours. The rest of
lines in (\ref{L10-p6}) contain the $\cO(p^6)$ corrections: 
the tree-level contributions from the $\cO(p^6)$ \chpt\ Lagrangian are given in the third  and fourth lines,
the term proportional to $(L_9^r+2 L_{10}^r)(\mu)$ in the 
fifth  line is the one-loop contribution of the
$\cO(p^4)$ \chpt\ Lagrangian, and the function 
$G_{2L}(\mu,s\!=\!0)$ in the last line,
 which does not depend on any LEC,
contains the proper two-loop contributions.

In Eq.~(\ref{C87-p6}) the tree-level contribution is given by $C_{87}^r(\mu)$, whereas the term proportional to $L_9^r(\mu)$ is a one-loop correction, which is suppressed by one power of $1/N_C$, and the two-loop contributions are contained in $G'_{2L}(\mu,s)\equiv \frac{d}{ds}G_{2L}(\mu,s)$.
The derivative operation, when acting over the one-loop contribution to $\Pi(s)$,
generates the terms proportional to inverse powers of the pion and kaon masses in the second line.
For simplicity, we omit the explicit analytic forms of
$G_{2L}(\mu)$ and $G'_{2L}(\mu)$, which are very lengthy and not too enlightening; these two functions contain
a $1/N_C^2$ suppression factor with respect to $L_{10}^r(\mu)$ and $C_{87}^r(\mu)$.

\section{Determination of Effective Couplings}

%%%%%%%%%%%%%%%%%%%%%%%%%%%FIGURE%%%%%%%%%%%%%%%%%%%%%%%%%%%%%%%%
%\begin{figure}[htb]
%\begin{center}
%\includegraphics[width=8cm]{ALEPHspecfun.eps}
%\end{center}
%\caption{V-A spectral function $\rho(s)=\frac{1}{\pi}\mbox{Im}
%\Pi^{0+1}_{ud,V-A}(s)$ measured by ALEPH \cite{ALEPH05}.}
%\label{fig:VmenosA}
%\end{figure}
%%%%%%%%%%%%%%%%%%%%%%%%%%%FIGURE%%%%%%%%%%%%%%%%%%%%%%%%%%%%%%%%

We will use the 2005 ALEPH data on semileptonic $\tau$ decays \cite{ALEPH05}, which provides the most recent and precise measurement of the $V-A$ spectral function. The effective chiral couplings can be directly extracted from the following integrals over the
hadronic spectrum:
\ba
\label{eq:defL10}
-8 \, L_{10}^{\rm eff}&\equiv& \overline{\Pi}(0)\, =\,
\frac{1}{\pi} \int^{s_0}_{s_{th}}
\, \frac{{\rm d} s}{s} \, \mathrm{Im} \,\Pi(s) \,
 , \\
\label{eq:defC87}
16 \, C_{87}^{ \rm eff} &\equiv&  \overline{\Pi}\,{}'(0)\, =\,
\frac{1}{\pi} \int^{s_0}_{s_{th}}
\, \frac{{\rm d} s}{s^2} \, {\rm Im} \,\Pi(s)\, .
\ea
These relations are exactly satisfied at $s_0\to\infty$. At finite values of $s_0$,
they assume that the OPE approximates well the correlator $\Pi(s)$ over the entire complex circle 
\footnote{Or equivalently these relations assume that the integrals on the real axis from $s_0$ to infinite are negligible, what is expected to be true only for high enough values of $s_0$ and for
accidental ``duality points''.} $|s|=s_0$. The OPE is expected to be a valid approximation for
high-enough values of $s_0$ and away from the real axis. While the kinematics of $\tau$ decay
restrict the upper limit of integration to the range $s_0\le m_\tau^2$,
the main source of theoretical uncertainty in the contour integration originates in the
region close to the point $s=s_0$ in the real axis.
Studying the sensitivity to $s_0$ of the integrals (\ref{eq:defL10}) and (\ref{eq:defC87}),
one can test validity of the OPE and assess the size of the associated systematic errors.

%%%%%%%%%%%%%%%%%%% FIGURE %%%%%%%%%%%%%%%%%%%%%
\begin{figure}[thb]
\vfill
\centerline{
\begin{minipage}[t]{.3\linewidth}\centering
\centerline{\includegraphics[width=9cm]{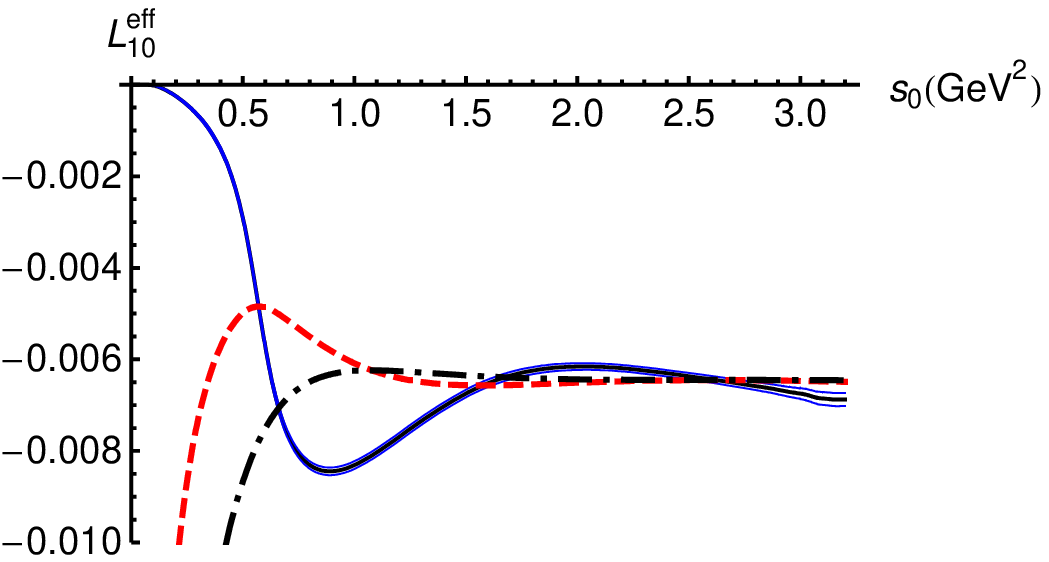}}
\end{minipage}
}
\vfill
\caption{Determinations of $L_{10}^{\rm eff}$ at different values of $s_0$.
The continuous lines show the results obtained from Eq.~(\ref{eq:defL10}).
The modified expressions in Eqs.~(\ref{eq:L10-ModSR1}) and (\ref{eq:L10-ModSR2}) give rise to the
dashed and dot-dashed lines, respectively. For clarity, we do not include their corresponding error bands.}
\label{fig:L10}
\end{figure}
%%%%%%%%%%%%%% END FIGURE %%%%%%%%%%%%%%%%%%%%%%

In Fig. \ref{fig:L10}, we plot the value of $L_{10}^{\rm eff}$ obtained from Eq.~(\ref{eq:defL10})
for different values of $s_0$. The band between the continuous lines shows the corresponding
experimental uncertainties (at one sigma). As expected, the result is far from an horizontal line
at low values of $s_0$, where the applicability of the OPE is suspect. The oscillatory behaviour
stabilises quite fast reaching a rather stable and flat result at
values of $s_0$ between 2 and 3 $\mathrm{GeV}^2$. The weight factor $1/s$ decreases the impact of the high-energy region,
minimising the size of quark-hadron duality violations around $s_0$.
This integral appears then to be much better behaved than the corresponding FESRs with $s^n$ ($n\ge0$) weights.

There are several possible strategies to estimate the central value for $L_{10}^{\rm eff}$
and the unavoidable theoretical uncertainties.
One is to give the predictions fixing $s_0$ at the so-called ``duality points'',
where the first and second WSRs happen to be satisfied.
Owing to the oscillatory behaviour of the WSRs results, this happens at two different values
of $s_0$. At the highest ``duality point'', which is obviously the more reliable, we obtain
$L_{10}^{\rm eff} = -(6.45 \pm 0.09) \cdot 10^{-3}$, where the quoted error only includes the experimental uncertainty.
Being very conservative, one could also take into account the first ``duality point'';
performing a weighted average of both results,  we get
$L_{10}^{\rm eff} = -(6.50\pm 0.13) \cdot 10^{-3}$, where the uncertainty covers the values obtained at the two
``duality points''.

Assuming that the integral \eqref{eq:defL10} oscillates around
his asymptotic value with decreasing oscillations, one can get another
estimate performing an average between the maxima and minima of the successive oscillations.
This procedure gives a value $L_{10}^{\rm eff} = -(6.5\pm 0.2) \cdot 10^{-3}$,
that is perfectly compatible with the previous results based on the ``duality points''.
Our last method of estimating the quark-hadron duality violation
uses appropriate oscillating functions defined in \cite{GON07}
which mimic the real quark-hadron oscillations above the data.
These functions are defined such that they match the data at approximately
3 $\mathrm{GeV}^2$, go to zero with decreasing oscillations and
satisfy the first and second WSRs. We find in this way
$L_{10}^{\rm eff} = -(6.50\pm 0.12) \cdot 10^{-3}$,
where the error spans the range generated by the different functions used.
This result agrees well with our previous estimates.

We can take advantage of the WSRs to construct modified sum rules with weight
factors proportional to $(1-s/s_0)$, in order to suppress numerically the role
of the suspect region around $s\sim s_0$ \cite{LDP:92}:
\ba
\label{eq:L10-ModSR1}
-8 \, L_{10}^{\rm eff}& = &
%%%\overline{\Pi}(0)\, =\,
\frac{1}{\pi} \int^{s_0}_{s_{th}}\, \frac{{\rm d} s}{s} \,\left(1-\frac{s}{s_0}\right)
\, \mathrm{Im} \,\Pi(s)
\, +\, \Delta_1(s_0)\, ,
%%%\frac{1}{s_0}\,\left(2 f_\pi^2 + C^{V-A}_2\right)\, ,
\nonumber \\  \\ [10pt] \label{eq:L10-ModSR2}
& = &
\frac{1}{\pi} \int^{s_0}_{s_{th}}\, \frac{{\rm d} s}{s} \,\left(1-\frac{s}{s_0}\right)^2
\, \mathrm{Im} \,\Pi(s) \nonumber \\
\, &+&\, 2 \Delta_1(s_0)\, -\,  \Delta_2(s_0)\, .\quad
%%%\frac{2}{s_0}\,\left(2 f_\pi^2 + C^{V-A}_2\right)
%%%\, -\, \frac{1}{s_0^2}\,\left(2 f_\pi^2 m_\pi^2- C^{V-A}_4\right)\, .
\ea
The factors $\Delta_1(s_0) = \left(2 f_\pi^2 + C^{V-A}_2\right)/s_0$ and
$\Delta_2(s_0) = \left(2 f_\pi^2 m_\pi^2- C^{V-A}_4\right)/s_0^2$ are small corrections dominated by the $f_\pi^2$ term, since
$C^{V-A}_{2,4}$ vanish in the chiral limit. The sum rule (\ref{eq:L10-ModSR2}) has been previously
used in refs.~\cite{DS07,DS04}.

The dashed and dot-dashed lines in Fig.~\ref{fig:L10} show the results obtained from
Eqs.~(\ref{eq:L10-ModSR1}) and (\ref{eq:L10-ModSR2}), respectively. As already found in refs.~\cite{DS07,DS04},
the modified weight factors minimise the theoretical uncertainties in a very sizeable way, giving rise
to very stable results over a quite wide range of $s_0$ values. One gets then
$L_{10}^{\rm eff} = -(6.51\pm 0.06) \cdot 10^{-3}$ using Eq.~(\ref{eq:L10-ModSR1}), and $L_{10}^{\rm eff} = -(6.45\pm 0.06) \cdot 10^{-3}$ from Eq.~(\ref{eq:L10-ModSR2}).

Taking into account all the previous discussion,
we quote as our final result:
\be
\label{L10eff}
L_{10}^{\rm eff} = -(6.48\pm 0.06) \cdot 10^{-3} \, .
\ee
%

%%%%%%%%%%%%%%%%%%% FIGURE %%%%%%%%%%%%%%%%%%%%%
\begin{figure}[thb]
\vfill
\centerline{
\begin{minipage}[t]{.3\linewidth}\centering
\centerline{\includegraphics[width=9cm]{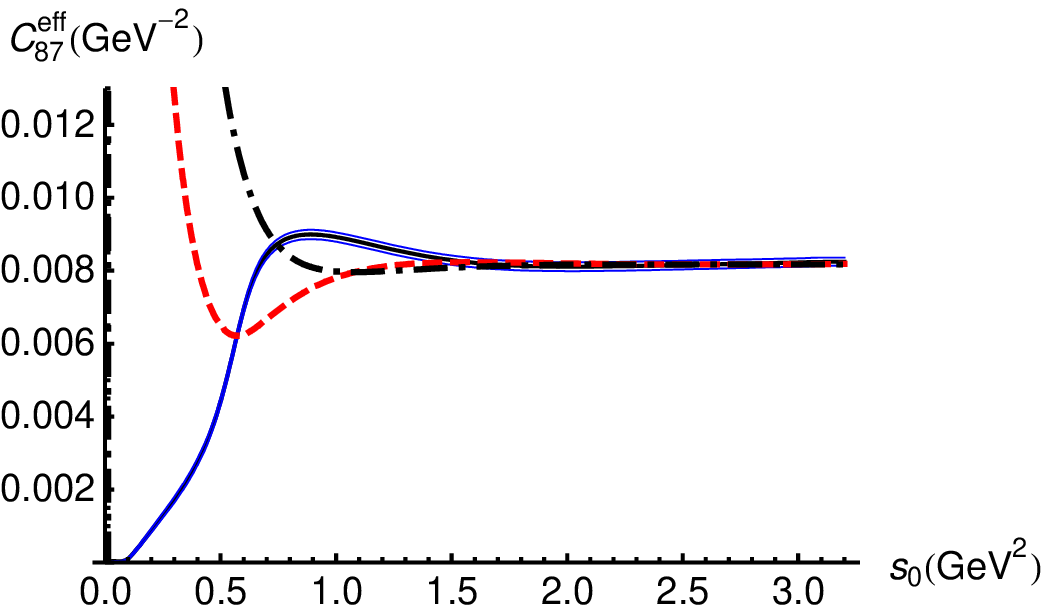}}
\end{minipage}
}
\vfill
\caption{Determinations of $C_{87}^{\rm eff}$ at different values of $s_0$.
The continuous lines show the results obtained from Eq.~(\ref{eq:defC87}).
The modified expressions in Eqs.~(\ref{eq:C87-ModSR1}) and (\ref{eq:C87-ModSR2}) give rise to the
dashed and dot-dashed lines, respectively. For clarity, we do not include their corresponding error bands.}
\label{fig:C87}
\end{figure}
%%%%%%%%%%%%%% END FIGURE %%%%%%%%%%%%%%%%%%%%%%

We have made a completely analogous analysis to determine the effective coupling $C_{87}^{\rm eff}$.
The results are shown in Fig.~\ref{fig:C87}. The continuous lines, obtained from Eq.~(\ref{eq:defC87}),
are much more stable than the corresponding results for $L_{10}^{\rm eff}$,
owing to the $1/s^2$ factor in the integrand.
The discontinuous and dotted lines correspond to the results obtained from the modified sum rules:
\ba
\label{eq:C87-ModSR1}
16 \, C_{87}^{\rm eff}& = &
\frac{1}{\pi} \int^{s_0}_{s_{th}}\, \frac{{\rm d} s}{s^2} \,\left(1-\frac{s^2}{s_0^2}\right)
\, \mathrm{Im} \,\Pi(s)
\, +\, \frac{\Delta_1}{s_0}\, ,
\\[10pt] \label{eq:C87-ModSR2}
& = &
\frac{1}{\pi} \int^{s_0}_{s_{th}}\, \frac{{\rm d} s}{s^2} \,\left(1-\frac{s}{s_0}\right)^2
\left(1+2\frac{s}{s_0}\right)\, \mathrm{Im} \,\Pi(s)
\, \nonumber \\
&+& \, \frac{3\Delta_1 -2 \Delta_2}{s_0}\, . \quad\quad
\ea
The agreement among the different estimates is quite remarkable.
We quote as our final conservative result,
\be
\label{C87eff}
C_{87}^{\rm eff} = (8.18\pm 0.14) \cdot 10^{-3} \, {\rm GeV}^{-2} \, .
\ee

\section{Determination of $L_{10}^r$ and $C_{87}^r$ }
\label{determination}

The \chpt\ coupling $L_{10}^{r}(\mu)$ can be obtained from $L_{10}^{\rm eff}$,
using the relation (\ref{L10-p6}). At $\cO(p^4)$ the determination is straightforward,
since one only needs to subtract from $L_{10}^{\rm eff}$ the term
$\left[1- \log{(\mu^2/m_\pi^2)}
+ \frac{1}{3}\log{(m_K^2/m_\pi^2)} \right]/(128\pi^2)$.
Taking $\mu=M_\rho$ as the reference value for the \chpt\ renormalization scale, one gets
\be
\label{valL10p4}
L_{10}^r(M_\rho) = -(5.22 \pm 0.06) \cdot 10^{-3}  \, .
\ee

At order $p^6$, the numerical relation is more subtle because it gets small corrections from other LECs.
It is useful to classify the $\cO(p^6)$ contributions through their ordering within the $1/N_C$ expansion.
The tree-level term $4 m_\pi^2 (C_{61}^r - C_{12}^r- C_{80}^r)(M_\rho)$, which is the only $\cO(p^6)$ correction
in the large--$N_C$ limit, is numerically small because it appears suppressed by a factor $m_\pi^2$.
The three relevant couplings have been determined phenomenologically with a moderate accuracy:
$C_{61}^r(M_\rho)=(1.24 \pm 0.44) \cdot 10^{-3}\:\mathrm{GeV}^{-2}$ \cite{KM06}
(from $\Pi^{(0+1)}_{ud,V}(0)-\Pi^{(0+1)}_{us,V}(0)$),
$C_{12}^r(M_\rho)=(0.4\pm 6.3) \cdot 10^{-5}\:\mathrm{GeV}^{-2}$ \cite{JOP04}
(from the $K\pi$ scalar form factor)
and
$C_{80}^r(M_\rho)=(2.1 \pm 0.5) \cdot 10^{-3}\:\mathrm{GeV}^{-2}$ \cite{UP08}
(from $a_1/K_1$ mass and width differences). These determinations agree reasonably well
with published meson-exchange estimates \cite{CEE05,ABT00} and lead to a total contribution
$4 m_\pi^2 (C_{61}^r - C_{12}^r- C_{80}^r)(M_\rho) = -(6.7\pm 5.2)\cdot 10^{-5}$. The scale dependence  
of this combination of $\cO(p^6)$  couplings \cite{p6}  
between $\mu= 0.6$ GeV and $\mu=1.1$ GeV  is within  its quoted 
uncertainty.

At  NLO in $1/N_C$ we need to consider the tree-level contribution proportional to
the combination of LECs $(C_{62}^r - C_{13}^r- C_{81}^r)(M_\rho)$. We are not aware of any published
estimate of these $1/N_C$ suppressed couplings, beyond the trivial statement that they don't get
any tree-level contribution from resonance exchange \cite{CEE05}.
We will adopt the conservative range
$|C_{62}^r - C_{13}^r- C_{81}^r|(M_\rho) \le |C_{61}^r - C_{12}^r- C_{80}^r|(M_\rho)/3$,
%%% assigning a 100\% uncertainty to this bold estimate; this gives a contribution
which gives a contribution
$4 (2 m_K^2+m_\pi^2)(C_{62}^r - C_{13}^r- C_{81}^r)(M_\rho) =
(0.0\pm 5.8)\cdot 10^{-4}$.
The scale dependence between $\mu= 0.6$ GeV and $\mu=1.1$ GeV
of this combination of $\cO(p^6)$  couplings \cite{p6}  
 is within its  quoted uncertainty.
The uncertainty on this term will dominate our final error on the $L_{10}^r(M_\rho)$ determination.
At the same NLO in $1/N_C$, there is also a one-loop correction proportional to $L_{9}^r(M_\rho)$;
using the $\cO(p^6)$ determination $L_9^r(M_\rho)=(5.93\pm 0.43)\cdot 10^{-3}$ \cite{BT02}, this
contribution can be estimated to be
$2 ( 2 \mu_\pi + \mu_K) \, L_9^r(M_\rho)
= - ( 1.56\pm 0.11)\cdot 10^{-3}$.
Finally, the $1/N_C^2$ suppressed two-loop function which collects the non-analytic contributions takes the value
$G_{2L}(M_\rho) = -0.524 \cdot 10^{-3}$,
one order of magnitude smaller than $L_{10}^{\rm eff}$,
but still eight
 times larger than the uncertainty quoted for $L_{10}^{\rm eff}$
in \eqref{L10eff}. Taking all these contributions into account,
we finally get the wanted $\cO(p^6)$ result:
\ba
\label{valL10p6}
L_{10}^r(M_\rho) &=&
-(4.06 \pm 0.04_{L_{10}^{\mathrm{eff}}}\pm 0.39_{\,\mathrm{LECs}})
\cdot 10^{-3} \nonumber \\
&=&  -(4.06 \pm 0.39) \cdot 10^{-3} \, ,
\ea
where the uncertainty has been split into its two main components.
The final error is completely dominated by our ignorance on the
$1/N_C$ suppressed LECs of $\cO(p^6)$.

The determination of $C_{87}^r$ from $C_{87}^{\rm eff}$ does not involve any unknown LEC.
The relation (\ref{C87-p6}) contains a one-loop correction
of size $-(3.15\pm 0.13)\cdot 10^{-3}$, which only depends on $L_9^r(M_\rho)$ and the
pion and kaon masses,
and small non-analytic two-loop contributions collected in the
term $G'_{2L}(M_\rho) = -0.277\cdot 10^{-3}\:\mathrm{GeV}^{-2}$.
In spite of its $1/N_C$ suppression, the one-loop correction is
very sizeable, decreasing the final value of the $\cO(p^6)$ LEC:
\be
\label{valC87}
C_{87}^r(M_\rho) = (4.89 \pm 0.19) \cdot 10^{-3} \:\mathrm{GeV}^{-2}\, .
\ee

\section{SU(2)  \chpt}

Up to now, we have discussed the LECs of the usual SU(3) \chpt.
It turns useful to consider also the effective low-energy theory with only two flavours of light quarks.
In some cases, this allows to perform high-accuracy phenomenological determinations of the corresponding LECs
at  NLO. Moreover, recent lattice calculations with two dynamical quarks are already able to obtain the
SU(2) LECs with sufficient accuracy and this is an important check for them.
 
In SU(2) \chpt, there are ten LECs, $l_{i=1,..7}$ and $h_{1,2,3}$, at $\cO(p^4)$ (NLO) \cite{GL84}.
Using the $\cO(p^6)$ relation between $l_5^r(\mu)$  and $L_{10}^r(\mu)$,
recently obtained in ref.~\cite{GHI07},
and  the definition of the invariant couplings $\overline l_i$ adopted in \cite{GL84}, we get
\ba
\overline l_5 &=& - 192 \pi^2 \, L_{10}^{\rm eff}
+ 1 + \log{\left( \frac{m_K}{\hat m_K} \right)}
\nonumber \\
&+& 768 \,  \pi^2 \, m_\pi^2 (C_{61}^r
+ C_{62}^r - C_{12}^r- C_{13}^r - C_{80}^r - C_{81}^r) (\mu)
\nonumber \\[7pt]
&+& 1536 \, \pi^2 (m_K^2 - \hat m_K^2)
( C_{62}^r -  C_{13}^r - C_{81}^r) (\mu)   \nonumber \\[7pt]
&-& 384\, \pi^2 (2\mu_\pi + \mu_K - \hat \mu_K)
 ( L_9^r + 2 L_{10}^r)(\mu) \nonumber \\
&-& x_K \left[ -\frac{67}{48} + \frac{21}{16} \rho_1 +
\frac{5}{8} \log \left(\frac{4}{3}\right)
-\frac{17}{4} \log{\left( \frac{\mu^2}{\hat m_K^2} \right)}
\right. \nonumber \\ &+& 
\left. 
\frac{3}{4} \log^2{ \left( \frac{\mu^2}{\hat m_K^2} \right)}
\right]
%+\frac{17}{4} x_K \log \left( \frac{\mu^2}{\hat m_K^2} \right)
%&-& \frac{3}{4} x_K \log^2 \left( \frac{\mu^2}{\hat m_K^2} \right)
+ 192 \,\pi^2 \, G_{2L}(\mu) \, +\, \cO(p^8)\, ,
\ea
where $\hat m_K^2 = m_K^2 - m_\pi^2/2$ is the kaon mass squared
in the limit $m_u=m_d=0$, 
$x_K= \hat m_K^2/(16 \pi^2 f_\pi^2)$, 
$\hat \mu_K= \hat m_K^2 \log(\hat m_K/\mu)/(16 \pi^2 f_\pi^2)$
and $\rho_1\simeq 1.41602$.

The first line contains the $\cO(p^4)$ contributions; the determination of
$\overline l_5$ at this order is then straightforward. The full $\cO(p^6)$ result,
with the different tree-level, one-loop and two-loop corrections,
is given in the other lines. Following the same procedure as in the SU(3) case, we get the results
\be
\overline l_5 \, =\,\left\{
\begin{array}{ccc}
 13.30 \pm 0.11 \, , & \; & \cO(p^4),\\[7pt]
 12.24 \pm 0.21 \, , & \; & \cO(p^6).
\end{array}\right.
\ee

\section{Summary}

Using the most recent hadronic $\tau$-decay data \cite{ALEPH05} on the $V-A$ spectral function,
and general properties of QCD such as analyticity, the OPE and \chpt,
we have determined very accurately the chiral LECs $L_{10}^r(M_\rho)$
and $C_{87}^r(M_\rho)$. Performing an $\cO(p^4)$ analysis, we obtain
\be\label{eq:res-p4}
L_{10}^r(M_\rho)=-(5.22 \pm 0.06) \cdot 10^{-3} \, ,
\ee
while a more elaborate study, including the $\cO(p^6)$ \chpt\ corrections
provides the values:
\ba\label{eq:res1-p6}
L_{10}^r(M_\rho)  &=&
-(4.06 \pm 0.04_{L_{10}^{\mathrm{eff}}}\pm 0.39_{\,\mathrm{LECs}}) \cdot 10^{-3} \nonumber \\
&=& -(4.06 \pm 0.39) \cdot 10^{-3} \, ,
\ea
and
\be \label{eq:res2-p6}
C_{87}^r(M_\rho)  = (4.89 \pm 0.19) \cdot 10^{-3} \: {\rm GeV}^{-2}\, .
\ee
Our error estimate includes a careful analysis of the theoretical uncertainties associated
with the use of the OPE in the dangerous region close to the physical cut. Moreover, in \eqref{eq:res1-p6}
we have explicitly separated the error into its two main components, showing that our present ignorance
on the $1/N_C$ suppressed LECs dominates the final uncertainty of the $L_{10}^r(M_\rho)$
determination at $\cO(p^6)$.

Several determinations of $L_{10}$ have been performed before \cite{DHG98,NAR01,DS04},
using the older 1998 ALEPH data \cite{ALEPH98,ALEPH97}.
In ref~\cite{DHG98} the result $L_{10}^r(M_\rho)= - (5.13 \pm 0.19) \cdot 10^{-3}$
was obtained to $\cO(p^4)$, through a simultaneous fit of this parameter and the OPE corrections
of dimensions six and eight to several spectral moments of the hadronic distribution.
This determination is in good agreement with our $\cO(p^4)$ result \eqref{eq:res-p4}.
Our quoted uncertainty has an smaller experimental contribution and includes a better assessment of the theoretical uncertainties.
The value $L_{10}^{\rm eff}=(-5.8\pm0.2)\cdot10^{-3}$ (3.2 $\sigma$ smaller than ours) was extracted from $\tau$ data in ref.~\cite{NAR01} using the first ``duality point'' of the WSRs. The difference comes from underestimated theoretical
uncertainties in this reference, as can be easily seen by choosing instead the second duality point or varying slightly the value of the first duality point. In fact the same reference \cite{NAR01} (see Eq. (10) therein) presents also a different estimate of $L_{10}^{\rm eff}$ that is in very good agreement with our result.
In ref.~\cite{DS04} both $L_{10}^{\rm eff}$ and $C_{87}^{\rm eff}$ were determined,
in good agreement with our findings which use the most recent 2005 data. An updated value
of $L_{10}^{\rm eff}$, using the 2005 data, has also been given in ref.~\cite{DS07}.

Our determinations of $L_{10}^r(\mu)$ and $C_{87}^r(\mu)$ at $\mu=M_\rho$ agree within errors
with the large--$N_C$ estimates based on lowest-meson dominance \cite{KN01,CEE04,ABT00,PI02}:
\ba
L_{10} & = & -\frac{F_V^2}{4 M_V^2}\, +\,\frac{F_A^2}{4 M_A^2}\,\approx\,  -\frac{3 f_\pi^2}{8 M_V^2}
\,\approx\, -5.4\cdot 10^{-3}\, , \nonumber \\
\\
C_{87} & = & \frac{F_V^2}{8 M_V^4}\, -\,\frac{F_A^2}{8 M_A^4}\,\approx\,\frac{7 f_\pi^2}{32 M_V^4}
\,\approx\, 5.3\cdot 10^{-3}\,\mathrm{GeV}^{-2} \, .
\nonumber \\
\ea
Eq. (\ref{eq:res2-p6}) is also in good
agreement with the result of ref. \cite{MP08} for $C_{87}$ based on 
Pad\'e Approximants. These predictions, however, are unable to fix the scale dependence which is of higher-order in $1/N_C$.
More recently, the resonance chiral theory Lagrangian \cite{CEE04,EGPdR89}
has been used to analyse the correlator $\Pi(s)$ at NLO order in the $1/N_C$ expansion \cite{PRS08}.
Matching the effective field theory description with the short-distance QCD behaviour, the two LECs
are determined, keeping full control of their $\mu$ dependence.
The theoretically predicted values $L_{10}^r(M_\rho) = -(4.4 \pm 0.9) \cdot 10^{-3}$
and $C_{87}^r(M_\rho)=(3.6 \pm 1.3) \cdot 10^{-3}$  GeV$^{-2}$ \cite{PRS08}
are in perfect agreement with our determinations, although less precise. A recent lattice estimate \cite{SHI08} finds $L_{10}^r(M_\rho) = -(5.2 \pm 0.5) \cdot 10^{-3}$ at $\cO(p^4)$, which is also in good agreement with our
 $\cO(p^4)$ result in (\ref{eq:res-p4}).

A recent reanalysis of the decay $\pi^+ \to e^+ \nu \gamma$  \cite{UP08}, using new experimental data,
has provided quite accurate values for the combination of $\cO(p^4)$ LECs $L_9+L_{10}$.
To $\cO(p^4)$ one finds $L_9^r(M_\rho)+L_{10}^r(M_\rho)=(1.32\pm 0.14)\cdot 10^{-3}$, while the
$\cO(p^6)$ result $L_9^r(M_\rho)+L_{10}^r(M_\rho)=(1.44\pm 0.08) \cdot 10^{-3}$ is slightly more precise \cite{UP08}.
Combining these numbers with our results for $L_{10}^r(M_\rho)$, one obtains
\be
L_9^r(M_\rho) \, =\,\left\{
\begin{array}{ccc}
 (6.54 \pm 0.15) \cdot 10^{-3} \, , & \; & \cO(p^4),\\[7pt]
 (5.50 \pm 0.40) \cdot 10^{-3} \, , & \; & \cO(p^6),
\end{array}\right.
\ee
in perfect agreement with the $\cO(p^4)$ result $L_9^r(M_\rho) = (6.9 \pm 0.7) \cdot 10^{-3}$ of ref. \cite{ECK07} and the $\cO(p^6)$ result $L_9^r(M_\rho) = (5.93 \pm 0.43) \cdot 10^{-3}$ of ref. \cite{BT02}. This 
last comparison represents an indirect check 
(in fact the only possible one for the moment) 
of our $\cO(p^6)$ result for $L_{10}$.

We have also determined the corresponding LEC of $L_{10}$ in the SU(2) effective theory, both at LO and NLO:
\be\label{eq:res3}
\overline l_5 \, =\,\left\{
\begin{array}{ccc}
 13.30 \pm 0.11 \, , & \; & \cO(p^4),\\[7pt]
 12.24 \pm 0.21 \, , & \; & \cO(p^6).
\end{array}\right.
\ee
>From a phenomenological analysis of the radiative decay $\pi \to l \nu \gamma$ within SU(2) \chpt, the authors of ref. \cite{BT97} obtained
$\overline l_6 - \overline l_5 = 2.57 \pm 0.35$ at $\cO(p^4)$, and
$\overline l_6 - \overline l_5 = 2.98 \pm 0.33$ at $\cO(p^6)$.
Using these results and our determinations for $\overline l_5$ in \eqref{eq:res3}, one gets
\be
\overline l_6\, =\,\left\{ 
\begin{array}{ccc}
 15.87 \pm 0.37 \, , & \; & \cO(p^4),\\[7pt]
 15.22 \pm 0.39 \, , & \; & \cO(p^6).
\end{array}\right.
\ee

At $\cO(p^4)$ the comparison of these estimates of SU(2) LECs with previous results is straightforward, since they are proportional to the corresponding SU(3) couplings, that we have already discussed. Our determination of
 $\overline l_5$  is the first one obtained at $\cO(p^6)$, whereas for $\overline l_6$ ref. \cite{BCT98} finds $\overline l_6 = 16.0 \pm 0.5 \pm 0.7$, where the last error is purely theoretical, in good agreement with ours, although less precise.

\section*{Acknowledgements}

We would like to thank Hans Bijnens and Pere Talavera for providing information on the two-loop functions in ref.~\cite{ABT00}. We are also grateful to Jorge Portol\'es for his help and useful comments.
M.G.-A. is indebted to MICINN (Spain) for a FPU Fellowship.
This work has been supported in part by the EU %%%European Commission
RTN network FLAVIAnet [Contract No. MRTN-CT-2006-035482],
by MICINN, Spain %%%and  FEDER (EC)
[Grants FPA2007-60323 (M.G.-A., A.P), FPA2006-05294 (J.P.) and Consolider-Ingenio
2010 Programme CSD2007-00042  --CPAN--] and
 by Junta de Andaluc\'{\i}a (J.P.)
[Grants P05-FQM 101, P05-FQM 467 and P07-FQM 03048].


\begin{thebibliography}{99}

\bibitem{ALEPH05} S. Schael {\it et al.}
[ALEPH Collaboration], Phys. Rep. {\bf 421} (2005) 191.
%%CITATION = PRPLC,421,191;%%

\bibitem{ALEPH98} R. Barate {\it et al.},
[ALEPH Collaboration], Eur. Phys.  {\bf C 4} (1998) 409.
Eur.\ Phys.\ J.\  C {\bf 4} (1998) 409.
%%CITATION = EPHJA,C4,409;%%

\bibitem{ALEPH97} R. Barate {\it et al.},
[ALEPH Collaboration], Z. Phys.  {\bf C 76} (1997) 15.
%%CITATION =ZEPYA,C76,15;%%

\bibitem{OPAL99} K. Ackerstaff  {\it et al.},
[OPAL Collaboration], Eur. Phys. {\bf C 7} (1999) 571.
%%CITATION = EPHJA,C7,571;%%

\bibitem{CLEO95} T. Coan {\it et al.},
[CLEO Collaboration], Phys. Lett. {\bf B 356} (1995) 571.
%%CITATION = PHLTA,B356,571;%%

\bibitem{Strange_data}
B. Aubert {\it et al.}, [BaBar Collaboration],
Phys. Rev. {\bf D 76} (2007) 051104;
%%CITATION = PHRVA,D76,051104;%%
Phys. Rev. Lett. {\bf 100} (2008) 011801;
%%CITATION = PRLTA,100,011801;%%
K. Inami {\it et al.},
[Belle Collaboration], Phys. Lett. {\bf B 643} (2006) 5;
%%CITATION = PHLTA,B643,5;%%
M.~Fujikawa {\it et al.}  [Belle Collaboration],
arXiv:0805.3773 [hep-ex].
%%CITATION = ARXIV:0805.3773;%%

\bibitem{alphas}
E. Braaten, Phys. Rev. Lett. {\bf 60} (1988) 1606;
%%CITATION = PRLTA,60,1606;%%
Phys. Rev. {\bf D 39} (1989) 1458;
%%CITATION = PHRVA,D39,1458;%%
S. Narison and A. Pich, Phys. Lett.
{\bf B 211} (1988) 183;
%%CITATION = PHLTA,B211,183;%%
E. Braaten, S. Narison and A. Pich,
Nucl. Phys. {\bf B 373} (1992) 581;
%%CITATION = NUPHA,B373,581;%%
F. Le Diberder and A. Pich,
Phys. Lett. {\bf B 286} (1992) 147;
%%CITATION = PHLTA,B286,147;%%
A. Pich, Nucl. Phys. B (Proc. Suppl.)
{\bf 39 B,C} (1995) 326.
%%CITATION = NUPHZ,39BC,326;%%

\bibitem{LDP:92}
F. Le Diberder and A. Pich, Phys. Lett. {\bf B 289} (1992) 165.
%%CITATION = PHLTA,B289,165;%%

\bibitem{DHZ06} M. Davier, A. H\"ocker and H. Zhang,
Rev. Mod. Phys. {\bf 78} (2006) 1043;
%%CITATION = RMPHA,78,1043;%%
M. Davier {\it et al.}, %%%arXiv:0803.0979 [hep-ph].
Eur. Phys. J. {\bf C 56} (2008) 305.
%%CITATION = EPHJA,C56,305;%%

\bibitem{new08}
P.A. Baikov, K.G. Chetyrkin and J.H. K\"uhn, %%%arXiv:0801.1821 [hep-ph];
 Phys. Rev. Lett.  {\bf 101} (2008) 012002;
%%CITATION = PRLTA,101,012002;%%
M. Beneke and M. Jamin, JHEP {\bf 09} (2008) 044;
 %%% [arXiv:0806.3156 [hep-ph]].
%%CITATION = JHEPA,0809,044;%%
K.~Maltman and T. Yavin, arXiv:0807.0650 [hep-ph].
%%CITATION = ARXIV:0807.0650;%%

\bibitem{review}
A. Pich, Int. J. Mod. Phys. {\bf A 21}
(2006) 5652;
%%CITATION = IMPAE,A21,5652;%%
Nucl. Phys. B (Proc. Suppl.) {\bf 169} (2007) 393;
%%CITATION =  NUPHZ,169,393;%%
arXiv:0806.2793 [hep-ph].
%%CITATION =  ARXIV:0806.2793;%%

\bibitem{su3}
A. Pich and J. Prades,
JHEP {\bf 06} (1998) 013;
%%CITATION = JHEPA,9806,013;%%
Nucl. Phys. B (Proc. Suppl.)  {\bf 74} (1999) 309;
  %%CITATION = NUPHZ,74,309;%%
JHEP {\bf 10} (1999) 004;
%%CITATION = JHEPA,9910,004;%%
Nucl. Phys. B (Proc. Suppl.)  {\bf 86} (2000) 236;
  %%CITATION = NUPHZ,86,236;%%
J. Prades, Nucl. Phys. B (Proc. Suppl.)  {\bf 76} (1999) 341;
  %%CITATION = NUPHZ,76,341;%%
S. Chen, {\it et al.}
  Eur. Phys. J.  {\bf C 22} (2001) 31;
  %%CITATION = EPHJA,C22,31;%%
M. Davier, {\it et al.}
  Nucl. Phys. B (Proc. Suppl.)  {\bf 98} (2001) 319;
  %%CITATION = NUPHZ,98,319;%%

\bibitem{CKP98}
K.G. Chetyrkin, J.H. K\"uhn and A.A. Pivovarov,
Nucl. Phys. {\bf B 533} (1998) 473;
  %%CITATION = NUPHA,B533,473;%%
P.A. Baikov, K.G. Chetyrkin and J.H. K\"uhn,
Phys. Rev. Lett. {\bf 95} (2005) 012003.
  %%CITATION = PRLTA,95,012003;%%

\bibitem{KKP01}
J.G. K\"orner, F. Krajewski and A.A. Pivovarov,
Eur. Phys. J {\bf C 20} (2001) 259.
  %%CITATION = EPHJA,C20,259;%%

\bibitem{MALT}
K. Maltman, Phys. Rev. {\bf D 58} (1998) 093015;
%%CITATION =  PHRVA,D58,093015;%%
J. Kambor and K. Maltman, ibid. {\bf D 62}
(2000) 093023;
%%CITATION =  PHRVA,D62,093023;%%
K. Maltman and J. Kambor, 
ibid.  {\bf D 64} (2001) 093014;
%%CITATION =  PHRVA,D64,093014;%%
K. Maltman and C.E. Wolfe, Phys. Lett.
{\bf B 639} (2006) 286;
%%CITATION =  PHLTA,B639,286;%%
K. Maltman {\it et al.}, arXiv:0807.3195 [hep-ph].
%%CITATION = ARXIV:0807.3195;%%

\bibitem{Vus}
E. G\'amiz, M. Jamin, A. Pich, J. Prades,
and F. Schwab, JHEP {\bf 01} (2003) 060;
  %%CITATION = JHEPA,0301,060;%%
  Phys. Rev. Lett.  {\bf 94} (2005) 011803;
  %%CITATION = PRLTA,94,011803;%%
 Nucl. Phys. B (Proc. Suppl.)  {\bf 144} (2005) 59;
  %%CITATION = NUPHZ,144,59;%%
arXiv:hep-ph/0505122;
  %%CITATION = HEP-PH/0505122;%%
arXiv:hep-ph/0610246;
  %%CITATION = HEP-PH/0610246;%%
Nucl. Phys. B (Proc.\ Suppl.)  {\bf 169} (2007) 85;
  %%CITATION = NUPHZ,169,85;%%
PoS {\bf KAON} (2008) 008.
  %%CITATION = POSCI,KAON,008;%%

\bibitem{DG:94}
J. F. Donoghue and E. Golowich, Phys. Rev. {\bf D 49} (1994) 1513.
%%CITATION = PHRVA,D49,1513;%%

\bibitem{DHG98} M. Davier, A. H\"ocker, L. Girlanda,
and J. Stern, Phys. Rev.  {\bf D 58} (1998) 096014.
%%CITATION =  PHRVA,D58,096014;%%

\bibitem{NAR01} S.~Narison,
  Nucl.\ Phys.\  B {\bf 593} (2001) 3.
  %%CITATION = NUPHA,B593,3;%%


\bibitem{WSR} S. Weinberg, Phys. Rev. Lett. {\bf 18} (1967) 507.
  %%CITATION = PRLTA,18,507;%%

\bibitem{DS07}
C.A. Dom\'{\i}nguez and K. Schilcher,
  JHEP {\bf 01} (2007) 093;
  %%CITATION = JHEPA,0701,093;%%
  J. Bordes, C.A. Dom\'{\i}nguez, J. Pe\~narrocha and K. Schilcher,
  JHEP {\bf 02} (2006) 037.
  %%CITATION = JHEPA,0602,037;%%

\bibitem{CGM03}
 V. Cirigliano, E. Golowich and K. Maltman,
  Phys.\ Rev.\  {\bf D 68} (2003) 054013.
  %%CITATION = PHRVA,D68,054013;%%

\bibitem{Q7Q8}
J.F. Donoghue and E. Golowich, Phys. Lett.
{\bf B 478} (2000) 172;
%%CITATION =  PHLTA,B478,172;%%
V. Cirigliano {\it et al.}, ibid. {\bf B 522} (2001) 245;
%%CITATION =  PHLTA,B522,245;%%
ibid. {\bf B 555} (2003) 71;
%%CITATION =  PHLTA,B555,71;%%
J. Bijnens, E. G\'amiz and J. Prades, JHEP {\bf 10}
(2001) 009;
%%CITATION = JHEPA,0110,009;%%
Nucl. Phys. B (Proc. Suppl.) {\bf 121} (2003) 195;
%%CITATION = NUPHZ,121,195;%%
S. Narison, Nucl. Phys. {\bf B 593} (2001) 3.
%%CITATION = NUPHA,B593,3;%%

\bibitem{WEI79}
S. Weinberg,  Physica {\bf A 96} (1979) 327.
  %%CITATION = PHYSA,A96,327;%%

\bibitem{GL84}
J. Gasser and H. Leutwyler,
Annals Phys.\  {\bf 158} (1984) 142.
  %%CITATION = APNYA,158,142;%%

\bibitem{GL85}
J. Gasser and H. Leutwyler,
 Nucl. Phys. {\bf B 250} (1985) 465.
%%CITATION = NUPHA,B250,465;%%

\bibitem{KdR94}
M. Knecht and E. de Rafael, 
Phys. Lett.  B {\bf 424} (1998) 335.
  %%CITATION = PHLTA,B424,335;%%

\bibitem{p6}
 J. Bijnens, G. Colangelo and G. Ecker,
Annals Phys.\  {\bf 280} (2000) 100;
  %%CITATION = APNYA,280,100;%%
  JHEP {\bf 02} (1999) 020;
  %%CITATION = JHEPA,9902,020;%%
H.W. Fearing and S. Scherer,
  Phys.\ Rev.\  {\bf D 53} (1996) 315.
  %%CITATION = PHRVA,D53,315;%%

\bibitem{ECK07}
 G. Ecker,   Acta Phys.\ Polon.\  {\bf B 38} (2007) 2753.
  %%CITATION = APPOA,B38,2753;%%

\bibitem{MOU97}
B. Moussallam, Nucl. Phys. {\bf B 504} (1997) 381.
%%CITATION = NUPHA,B504,381;%%

\bibitem{KN01}
M. Knecht and A. Nyffeler, Eur. Phys. J  {\bf C 21} (2001) 659.
%%CITATION = ZEPYA,C21,659;%%

\bibitem{RPP03}
P. Ruiz-Femen\'{\i}a, A. Pich and J. Portol\'es, JHEP {\bf 07} (2003) 003.
%%CITATION = JHEPA,0307,003;%%

\bibitem{BGL03}
J. Bijnens, E. G\'amiz, E. Lipartia and J. Prades,
  JHEP {\bf 04} (2003) 055.
%%CITATION = JHEPA,0304,055;%%

\bibitem{KM06}
K. Kampf and B. Moussallam,
Eur. Phys. J. {\bf C 47} (2006) 723;
%%CITATION = ZEPYA,C47,723;%%
S. D\"urr and J. Kambor, Phys. Rev.
{\bf D 61} (2000) 114025.
%%CITATION = PHRVA,D61,114025;%%

\bibitem{CEE05} V. Cirigliano {\it et al.},
JHEP {\bf 04} (2005) 006.
  %%CITATION = JHEPA,0504,006;%%

\bibitem{CEE04}
V. Cirigliano {\it el al.},
Nucl. Phys. {\bf B 753} (2006) 139;
%%CITATION =NUPHA,B753,139;%%
 Phys.\ Lett.\  B {\bf 596} (2004) 96.
  %%CITATION = PHLTA,B596,96;%%

\bibitem{RSP07} I. Rosell, J.J. Sanz-Cillero and A. Pich,
  JHEP {\bf 01} (2007) 039.
  %%CITATION = JHEPA,0701,039;%%

\bibitem{MP08}
P. Masjuan and S. Peris,
 Phys.  Lett.  {\bf B 663} (2008) 61;
  %%CITATION = PHLTA,B663,61;%%
  JHEP {\bf 05} (2007) 040.
  %%CITATION = JHEPA,0705,040;%%

\bibitem{PRS08}
A. Pich, I. Rosell and J.J. Sanz-Cillero,
 JHEP {\bf 07} (2008) 014. %%  [arXiv:0803.1567 [hep-ph]].
%%CITATION = JHEPA,0807,014;%%

 \bibitem{DS04}
 C.A. Dom\'{\i}nguez and K. Schilcher,
  Phys.\ Lett.\  {\bf B 581} (2004) 193;
  %%CITATION = PHLTA,B581,193;%%
  ibid. {\bf B 448} (1999) 93.
  %%CITATION = PHLTA,B448,93;%%

\bibitem{ABT00} G. Amor\'os, J. Bijnens and P. Talavera,
Nucl. Phys. {\bf B 568} (2000) 319.
%%CITATION = NUPHA,B568,319;%%

\bibitem{GON07} M. Gonz\'alez-Alonso,
Val\`encia Univ. Master Thesis (2007).

\bibitem{JOP04} M. Jamin, J.A. Oller and A. Pich, JHEP {\bf 02} (2004) 047.
%%CITATION = JHEPA,0402,047;%%

\bibitem{UP08}
R. Unterdorfer and H. Pichl,
  Eur.\ Phys.\ J.  {\bf C 55} (2008) 273.
  %%CITATION = EPHJA,C55,273;%%

\bibitem{BT02} J. Bijnens and P. Talavera,
JHEP {\bf 03} (2002) 046.
%%CITATION = JHEPA,0203,046;%%

\bibitem{GHI07}
 J. Gasser, C. Haefeli, M.A. Ivanov and M. Schmid,
  Phys.\ Lett.\  B {\bf 652} (2007) 21.
  %%CITATION = PHLTA,B652,21;%%

\bibitem{PI02} A. Pich, arXiv:hep-ph/0205030.
%%CITATION = HEP-PH/0205030;%%

\bibitem{EGPdR89}
G. Ecker, J. Gasser, A. Pich and E. de Rafael, Nucl. Phys.
{\bf B 321} (1989) 311;
%%CITATION = NUPHA,B321,311;%%
G. Ecker, J. Gasser, H. Leutwyler, A. Pich and E. de Rafael,
Phys. Lett. {\bf B 223} (1989) 425.
%%CITATION = PHLTA,B223,425;%%

\bibitem{SHI08}
E. Shintani {\it et al.} [JLQCD Collaboration],
arXiv: 0806.4222 [hep-lat].
%%CITATION = ARXIV:0806.4222;%%

\bibitem{BT97}
J. Bijnens and P. Talavera,
Nucl. Phys. B {\bf 489} (1997) 387.
%%CITATION = NUPHA,B489,387;%%

\bibitem{BCT98}
J.~Bijnens, G.~Colangelo and P.~Talavera,
JHEP {\bf 05} (1998) 014
%%CITATION = JHEPA,9805,014;%%

\end{thebibliography}
\end{document}